\documentclass[prb,superscriptaddress,twocolumn,preprintnumbers,a4,showpacs,floatfix]{revtex4}
\usepackage{amsmath, amssymb, psfrag, tabls, dcolumn, bm, color}
\usepackage[sort&compress]{natbib}
\usepackage[pdftex]{graphicx}
\usepackage{graphicx}

\newcommand\smallurl[1]{{\tiny \url{#1}}}

\newcommand{\be}{\begin{equation}}
\newcommand{\ee}{\end{equation}}
\newcommand{\bea}{\begin{equation*}}
\newcommand{\eea}{\end{equation*}}
\newcommand{\ba}{\begin{array}}
\newcommand{\ea}{\end{array}}
\newcommand{\beqa}{\begin{eqnarray}}
\newcommand{\eeqa}{\end{eqnarray}}
\newcommand{\beqaa}{\begin{eqnarray*}}
\newcommand{\eeqaa}{\end{eqnarray*}}

\newcommand{\matr}{\left( \begin{array}}
\newcommand{\ematr}{\end{array} \right)}

\newcommand{\rb}{\mbox{\boldmath $r$}}
\newcommand{\Rb}{\mbox{\boldmath $R$}}

\newcommand{\cb}{\mbox{\boldmath $c$}}

\newcommand{\nb}{\mbox{\boldmath $n$}}

\newcommand{\iv}{\mbox{$\hat{\imath}$}}
\newcommand{\jv}{\mbox{$\hat{\jmath}$}}

\newcommand{\lsim}{{\;\raise0.3ex\hbox{$<$\kern-0.75em\raise-1.1ex\hbox{$\sim$}}
\;}}
\newcommand{\gsim}{{\;\raise0.3ex\hbox{$>$\kern-0.75em\raise-1.1ex\hbox{$\sim$}}
\;}}

\usepackage{mdwlist}

\def\bcols{\begin{columns}}
\def\ecols{\end{columns}}
\def\bcol{\begin{column}}
\def\ecol{\end{column}}
\def\bit{\begin{itemize}}
\def\eit{\end{itemize}}
\def\ben#1{\begin{enumerate}[#1]}
\def\een{\end{enumerate}}
\def\colb#1{\begin{columns}\begin{column}{#1}}

\def\cole{\end{column}\end{columns}}

\usepackage{bm}

\def\sop{\mathcal{S}}
\def\sopn{\mathcal{S}^{\bm n}}
\def\dsopn{\hat{D}(\mathcal{S}^{\bm n})}

\begin{document}

\title{Approximate Modeling of Spherical Membranes}

\author{Pekka Koskinen\footnote{Corresponding author}}
\email[email:]{pekka.koskinen@iki.fi}
\address{NanoScience Center, Department of Physics, University of Jyv\"askyl\"a, 40014 Jyv\"askyl\"a, Finland}

\author{Oleg O. Kit}
\address{NanoScience Center, Department of Physics, University of Jyv\"askyl\"a, 40014 Jyv\"askyl\"a, Finland}





\begin{abstract}
Spherical symmetry is ubiquitous in nature. It's therefore unfortunate that simulation of spherical systems is so hard, and require complete spheres with millions of interacting particles. Here we introduce a method to model spherical systems, using revised periodic boundary conditions adapted to spherical symmetry. Method reduces computational costs by orders of magnitude, and is applicable for both solid and liquid membranes, provided the curvature is sufficiently small. We demonstrate the method by calculating the bending and Gaussian curvature moduli of single- and multi-layer graphene. The method works with any interaction (\emph{ab initio}, classical interactions), with any approach (molecular dynamics, Monte Carlo), and with applications ranging from science to engineering, from liquid to solid membranes, from bubbles to balloons.
\end{abstract}

\maketitle

\section{Introduction to modeling approach}

The problem in simulating spherical symmetry is topological: you cannot build a perfect sphere from identical blocks. The absence of such a building block has enforced expensive simulations with complete spheres---though usually spherical simulations are simply avoided. Overwhelming dilemmas like this are often considered so fundamental and frustrating that they restrain all attempts to seek for a practical solution.

Anyhow, avoiding spherical systems in our world is hard. Spherical shells surround us in a variety of forms: in balloons, in cell membranes inside our bodies, in bubbles in the sea, or in Earth's crust. The interaction of nanoparticles with cell membranes, for instance, is a topical question.\cite{monticelli_SM_09} Since cell membranes' curvature moduli \emph{determine} the very forms of red blood cells, for example, one can see why simulations should incorporate curvature effects.\cite{marsh_CPL_06,markvoort_JPCB_06} Another timely example is the foam of spherical bubbles in the sea, the bursting of which may play an important role on the so-called sea spray that produces spherical aerosols into the atmosphere.\cite{bird_nature_10,spiel_JGR_98}

Although liquid membranes are more abundant in nature, also man-made solid membranes have spherical symmetries, at least locally. Examples are fullerenes,\cite{kroto_nature_85} nanoballoons,\cite{leenaerts_APL_08} and especially graphene that contains intrinsic ripples even when suspended freely.\cite{meyer_nature_07,bao_nnano_09,shenoy_PRL_08} Curvature moduli of graphene are intimately related to these ripples, whether they are intrinsic or not,\cite{thompson-flagg_EPL_09,wang_PRB_09} and in a broad sense to elastic behavior of all honeycomb carbon, among graphene nanoribbons,\cite{shenoy_PRL_08} multilayer graphene,\cite{huang_PNAS_09} and carbon nanotubes.\cite{pastewka_PRB_09, malola_PRB_08b}

Conventionally spherical systems are treated in three ways. The first way is to simulate the system as a whole. Needless to say, this is expensive and often impossible.\cite{markvoort_JPCB_06} The second way is, should the system have some well-defined point-group symmetries, to use those symmetries for reducing computational costs. Most established codes have the ability to benefit from such symmetries; this has long been a standard procedure with molecules and clusters.\cite{atkins_book_00} Because the symmetry is exact, however, neither the curvature nor other geometrical parameters can be changed flexibly. The third way is to ignore curvature altogether and to use periodic boundary conditions (PBC) to simulate an infinitely large, flat membrane. Unfortunately, in nanoscience many systems fall between these two extremes: systems with huge number of particles, having no overall symmetry, but prominent curvature effects. At the moment a practical way to simulate such systems does not exist.

The periodic boundary conditions have been adapted, however, also to symmetries beyond translation. The first ideas came along chiral carbon nanotubes,\cite{white_PRB_93,popov_NJP_04} and those ideas have been used ever since; for reviews look at Refs.~\onlinecite{white_JPCB_05} and \onlinecite{gunlycke_PRB_08}. An important extension to general symmetries with exact treatment was done in Ref.~\onlinecite{dumitrica_JMPS_07}, which has enabled more flexibility.\cite{zhang_JCP_08,nikiforov_APL_10,zhang_PRL_10}. Later, Ref.~\onlinecite{koskinen_PRL_10} introduced revised periodic boundary conditions (RPBC), a simple formalism for general material distortions; this is the approach we shall use here, and it's illustrating to review it briefly.

In RPBC, the usual translation operations are replaced by general symmetry operations $\sopn$ that, in a quantum-mechanical language, leave the electronic potential invariant, or
\begin{equation}
\dsopn V(\rb)\equiv V(\mathcal{S}^{-{\bm n}}\rb)=V(\rb).
\end{equation}
The operation $\sopn$ is a succession of an abelian group of operations $\sop_i$, that is $\sopn=\sop_1^{n_1} \sop_2^{n_2} \cdots$. Then, by imposing periodicity ($\mathcal{S}_i^{M_i}=1$, $M_i$ integer), one finds that the Hamiltonian eigenstates $\psi_{a {\bm \kappa}}(\rb)$ at $\rb$ and at $\rb'=\mathcal{S}^{-\bm n} \rb$ differ only by a phase factor,
\begin{equation}
\hat{D}(\mathcal{S}^{\bm n})\psi_{a {\bm \kappa}}(\rb)=\psi_{a {\bm \kappa}}(\mathcal{S}^{-{\bm n}} \rb)=\exp(i{\bm \kappa}\cdot \nb)\psi_{a {\bm \kappa}}(\rb),
\label{eq:bloch}
\end{equation}
with inverse operation $\mathcal{S}^{-{\bm n}}$, band index $a$, and the reciprocal lattice vector ${\bm \kappa}$. Eq.(\ref{eq:bloch}) infers the familiar result: a single simulation cell---whatever its shape---is enough to describe the extended system as a whole. Revised PBC is hence similar to conventional PBC and differs only in the definitions of the symmetry operations. There are no other fundamental differences. As an illustrative example, the total energy with a classical pair potential is
\begin{equation}
E_\text{pair} = \frac{1}{2}\sum_{i,j=1}^N \sum_{\bm n}  U_{ij}(|\Rb_i-\sopn \Rb_j|),
\end{equation}
where $N$ is the particle count and $\nb$ runs over operations where particle $i$ at $\Rb_i$ still interacts with the periodic image of particle $j$ at $\sopn \Rb_j$. Forces are the negative gradients of this expression, as usual. Look at Ref.~\onlinecite{koskinen_PRL_10} for details of RPBC and Refs.~\onlinecite{malola_PRB_08b} and \onlinecite{koskinen_PRB_10} for examples of usage.

In this paper we use the RPBC, reviewed above, in an approximate way to introduce a trick for modeling spherical membranes. Adapting RPBC for spherical systems enables simulations with orders-of-magnitude reductions in computational costs. We shall apply the method to calculate graphene's mean and Gaussian curvature moduli, but first we proceed to discuss symmetry operations and their character.

\section{Sphericity as an Approximate Symmetry}

\begin{figure}[tb]
\includegraphics[width=5.5cm]{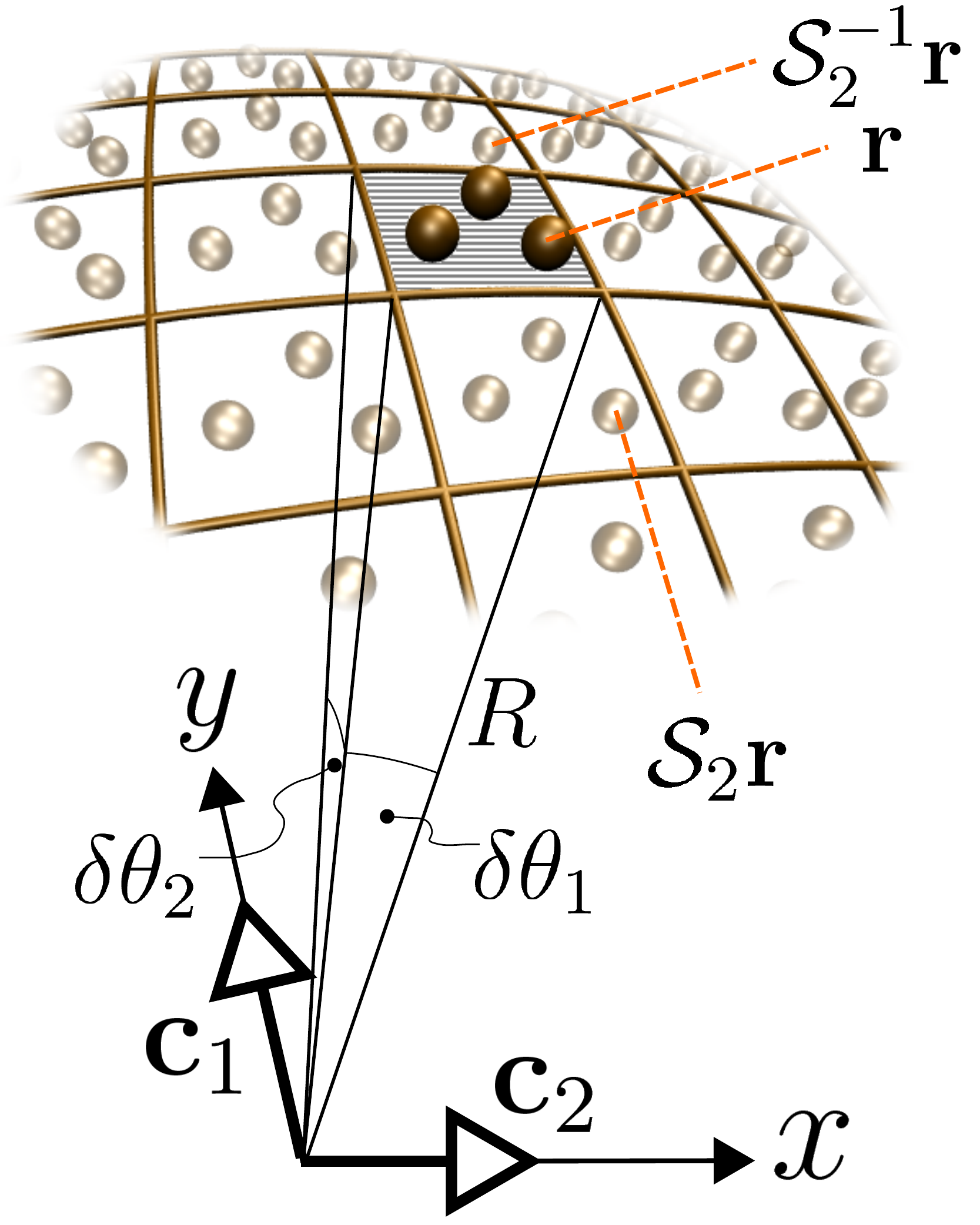}
\caption{(Color online) (a) Illustration of symmetry operations $\mathcal{S}_1$ and $\mathcal{S}_2$ for spherical symmetry. $\mathcal{S}_1$ is a rotation of angle $\delta \theta_1$ around $y$-axis and $\mathcal{S}_2$ is a rotation of angle $\delta \theta_2$ around $x$-axis; the angles $\delta \theta_i$ are small. (In general, $\cb_1$ can also be non-orthogonal to $\cb_2$ and $\delta \theta_1$ different from $\delta \theta_2$.)}
\label{fig:operations}
\end{figure}

Consider the square cone in Fig.~\ref{fig:operations}, regard the grid as fixed in space, and concentrate on the shaded region. If we rotate all particles an angle $\delta \theta_1$ around $y$-axis, or an angle $\delta \theta_2$ around $x$-axis, the geometry within the shaded region will remain approximately intact. This means that rotations $\mathcal{S}_1^{n_1} \rb=\mathcal{R}(n_1 \delta \theta_1 \cb_1)\rb$ and $\mathcal{S}_2^{n_2} \rb=\mathcal{R}(n_2 \delta \theta_2 \cb_2) \rb$ (with $\cb_1=\jv$, $\cb_2=\iv$ and operation $\mathcal{R}(\cb)$ as $|\cb|$-radian rotation around $\hat{\cb}$) leave the electronic potential $V(\rb)$ invariant near the shaded region: $\mathcal{S}_1$ and $\mathcal{S}_2$ are symmetry operations \emph{as far as the shaded region and its vicinity is concerned}. Two rotations around different axes do not commute in general, but if the rotation angles $\delta \theta_i$ are small $\mathcal{S}_i\rb\approx \rb + \delta \theta_i \cb_i \times \rb$, rotations do commute to linear order in $\delta \theta_i$'s, $[\mathcal{S}_1,\mathcal{S}_2]=\mathcal{O}(\delta \theta_i^2)$. Hence also the combined operation
\begin{equation}
(\mathcal{S}_1^{n_1}\mathcal{S}_2^{n_2}) \rb\approx \mathcal{R}(n_1\delta \theta_1\cb_1 + n_2\delta \theta_2 \cb_2)\rb \equiv \sopn \rb
\label{eq:rotation-operation}
\end{equation}
is an approximate symmetry operation, provided that $n_i$ are small enough. Eq.(\ref{eq:rotation-operation}) is basically all we need to fully employ the RPBC of Ref.~\onlinecite{koskinen_PRL_10}; we are, in principle, ready to go and to simulate any spherical membrane.

\section{Features due to approximation}

In practice, however, the approximate character of $\sopn$ raises questions that deserve some elaboration. First, as already mentioned, the formalism assumes periodic boundary conditions ($\sop_i^{M_i}=1$) which may seem questionable. Here we remind that similar PBCs are used also in regular bulk, with all three dimensions periodic in an intertwined fashion. (In two dimensions PBC represents topologically a toroid.) The bottom line is that periodicity is not a physical reality but a mere mathematical trick that works, and enables the application of revised Bloch's theorem in the first place.\cite{hansen_PRL_79,kratky_JCP_80} The integers $M_i$ are connected rather to ${\bm \kappa}$-point sampling than to physical reality.

Second, revised PBC does not need the ``unit cell'' concept. However, we shall call the square cone in Fig.~\ref{fig:operations}, extending from the origin to infinity and enclosing the shaded region, a unit or simulation cell because the concept is familiar and convenient in discussion. Otherwise, the mere expression for $\sopn$ in Eq.(\ref{eq:rotation-operation}) is enough to determine everything in the simulation.

Third, the claim is not to simulate a complete sphere, but rather to \emph{view the curvature as a local property}. The particles in the simulation cell see the closest environment curved---and only this is important. The simulation cell is the only cell we model, and distances and angles measured only from the simulation cell are meaningful. For example, the vicinity of particle at $\rb$ in Fig.\ref{fig:operations} exhibits curvature in bond angles and distances if one looks at particle's own periodic images at $\sop_2 \rb$ and $\sop_2^{-1}\rb$. Symmetry operations $\sopn$ that have $n_i$ large enough to rotate large angles ($n_i\sim (\pi/2)/\delta \theta_i$) should be excluded because the non-commutativity of $\mathcal{S}_i$'s would otherwise become significant.

Fourth, the radius or curvature $R$ in Fig.~\ref{fig:operations} is not a parameter in the simulation; radially particles can migrate wherever interactions drive them. The spherical form is only forced by the choice of symmetry operations and the parameters $\delta \theta_1$ and $\delta \theta_2$, and since the symmetry is discrete, the system needs to be neither continuously nor smoothly spherical.

Fifth, a natural limitation is to have enough empty space near the origin to avoid too close encounters between the particles.\cite{void} Membrane can be thick.

\section{Applying the method}

The validity of the method depends on the system and its interactions. As a principal rule, the radius of curvature $R$ should be much larger than the interaction ranges between the particles. If ranges are larger than the system size, especially if those interactions control morphology, one does better to model the complete system. The Coulomb interactions can play a role locally, within small length scales (size of the unit cell at most), but the long-ranged Coulomb interaction requires special care, perhaps some refinements (the unit cell better be neutral).\cite{electrostatics} Quantitative error due to the non-commutativity of $\mathcal{S}_i$'s can be estimated by first using the right-hand side of Eq.(\ref{eq:rotation-operation}) as $\sopn$, and then using the left-hand side of Eq.(\ref{eq:rotation-operation}) as $\sopn$ (changing the ordering of $\sop_1^{n_1}\sop_2^{n_2}$), and comparing the resulting energies.

Because liquid lacks long-range order, the method suits particularly well for liquid membranes, such as lipid bilayers. Their energetics can be described by the Helfrich Hamiltonian that gives membrane's elastic energy per unit area as~\cite{helfrich_NC_73}
\begin{equation}
g=2\kappa \left( \frac{1}{2}\left[ \frac{1}{R_1}+\frac{1}{R_2} \right] \right)^2 + \overline{\kappa} \frac{1}{R_1 R_2}.
\label{eq:helfrich}
\end{equation}
Here $\kappa$ is the mean curvature modulus (don't confuse with a ${\bm \kappa}$-point), $\overline{\kappa}$ is the Gaussian curvature modulus, and $R_1$ and $R_2$ are the principal radii of curvature. The liquid membrane doesn't need to be free-standing, however, because also solid support can be incorporated, either by external force fields or by fixed atoms. External radial forces can be also used for pressurization, mimicking the embedding of membrane in gaseous or liquid environments.

For solid membranes the situation is more complicated, because energy will come also from the internal strain $E_s$. If a flat, round sheet of radius $\rho$ is wrapped into a spherical segment, the energy will be $E_s \sim Eh\pi \rho^6/108 R^4$,\cite{Es} where $E$ is the Young's modulus of the material, $h$ is the membrane thickness, and $R$ is the radius of curvature; meanwhile the curvature-related energy is $E_c=g\cdot \pi \rho^2$. Hence, for a reasonable modeling of solid membranes using Eq.(\ref{eq:helfrich}), we need to have $E_s\ll E_c$, or
\begin{equation}
E_s/E_c \sim \frac{Eh \rho^4 R_\text{min}^{-2}}{108 \cdot (2\kappa + \bar{\kappa})}\ll 1,
\label{eq:criterion}
\end{equation}
which suggests a minimum radius of curvature $R_\text{min}$ for a given unit cell area. If this geometrical and material-dependent criterion should be violated, the simulation would be dominated by non-local stress fields. Since the method does not properly describe these fields, the treatment would become ill-defined.

The above problem is present when sphericity is forced on originally flat sheet. But defects, for example, can induce spontaneous curvature in solid membranes in which case $R_\text{min}$ can be smaller. The method provides a new tool to investigate phenomena such as rippling due to adsorption-induced pinching of the membrane.\cite{thompson-flagg_EPL_09} This method does not directly compete with any existing method, but instead it provides possibilities to do something new.

%
%
\section{Example: spherical graphene}

The spherical symmetry was implemented in the density-functional tight-binding code \texttt{hotbit}.\cite{koskinen_CMS_09,hotbit_wiki} The RPBC implementation has a negligible computational overhead as compared to translational symmetry,\cite{koskinen_PRL_10} and can be implemented just by a few lines of new code in any existing RPBC implementation. The code source is open and stands for inspection.

In this section we use the \texttt{hotbit} implementation to present one practical example. We calculate the curvature moduli of graphene, motivated by their relevance to present-day engineering with carbon nanostructures. For a sphere the radii of curvature are $R_1=R_2=R$, and Eq.(\ref{eq:helfrich}) gives $g=(2\kappa+\overline{\kappa})/R^2$; for a cylinder $R_1=R$, $R_2=\infty$, and $g=\kappa/(2R^2)$. Hence, by calculating the elastic energies for a cylinder and a sphere and varying $\delta \theta_i$'s (hence varying $R$) we obtain both $\kappa$ and $\overline{\kappa}$ directly.

\begin{figure}[tb]
\includegraphics[width=7.5cm]{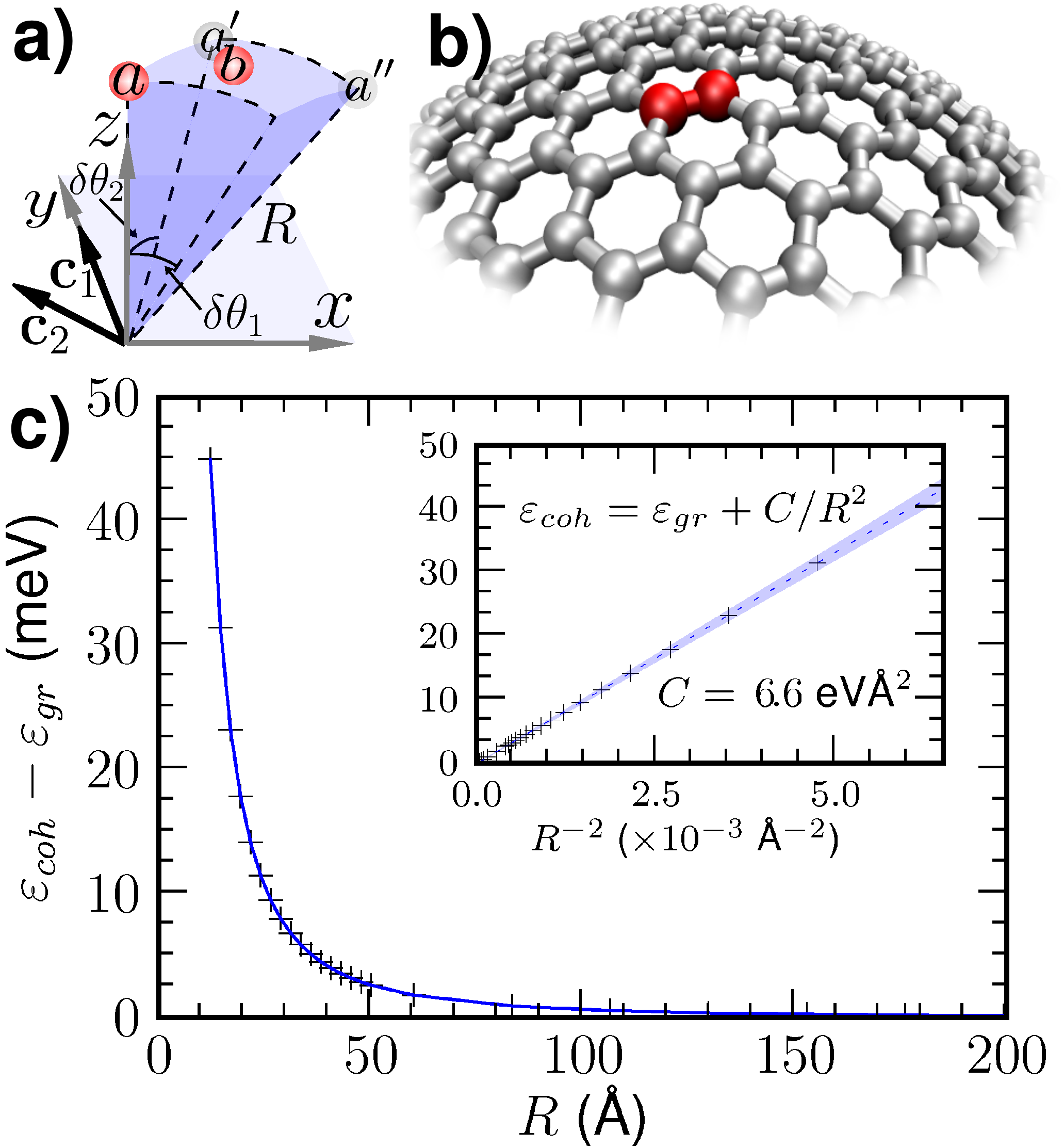}
\caption{(Color online) (a) Two-atom unit cell for spherical graphene, illustrating the symmetry operations: $\sop_1$ is a rotation of angle $\delta \theta_1$ around $\cb_1$ and $\sop_2$ is a rotation of angle $\delta \theta_2$ around $\cb_2$. (b) Few periodic images of atoms $a$ and $b$, shown for visualization purposes only. (c) Elastic energy per atom as a function of radius of curvature. Inset: fit to $R^{-2}$ behavior; the thin shaded fan is the error estimate due to approximations involved.}
\label{fig:graphene}
\end{figure}

Prior to simulating spherical graphene, we first calculated the mean curvature modulus of graphene, also applying revised PBC. Only now the symmetry operations, in a cylinder-like setup, were a rotation around $z$-axis ($\sop_1$) and translation in $z$-direction ($\sop_2$) with a $4$-atom unit cell (like a nanotube with enormous diameter); we won't discuss the cylinder setup further here.\cite{popov_NJP_04} The resulting cohesive energy depends on $R$ quantitatively like $R^{-2}$, as Eq.(\ref{eq:helfrich}) suggests, and the fitted value for $\kappa=1.61$~eV ($4.22$~eV\AA$^2$/atom) agrees with a density-functional reference value ($1.5$~eV)\cite{kudin_PRB_01} albeit is larger than an experimental reference value ($1.2$~eV).\cite{nicklow_PRB_72}

Returning to spherical graphene, Figs.~\ref{fig:graphene}a and \ref{fig:graphene}b show the two-atom unit cell of graphene. Unlike in Fig.\ref{fig:operations}, the unit cell is skewed with $\cb_1 = \jv$ and $\cb_2 = \cos (5\pi/6) \iv + \sin (5\pi/6) \jv$. The geometry was optimized with given $\delta \theta_i$'s, which were taken as $\delta \theta_i=2.5$~\AA$/R'$ when we wanted to investigate a radius of curvature that roughly equals $R'$.\cite{bitzek_PRL_06} All the radii of curvature we report, anyhow, are the optimized $R$ ($R\approx R'$ because curvature changes bond distances only slightly). In practice we found that structure optimizations require convergence criteria tighter than with translational cells, due to geometrical effects from small $\delta \theta_i$.\cite{fmax}  In quantum simulations ${\bm \kappa}$-points can be freely sampled ($\kappa_i \in[-\pi,\pi]$) because PBC is an approximation, just as with conventional Bloch's theorem; we used a $50\times 50$ ${\bm \kappa}$-point mesh.

Fig.\ref{fig:graphene}c shows our main result, graphene's cohesive energy as a function curvature---and represents the showcase of the new physics this method can unearth. Energy behaves clearly like $R^{-2}$, as suggested by Eq.(\ref{eq:helfrich}). The energy penalty $6.6$~eV\AA$^2R^{-2}$/atom, combined with previously calculated $\kappa$, yields $\overline{\kappa}=-0.70$~eV; we could not find this number in the literature. This result confirms graphene's beautiful elastic behavior up to high curvature---also for spherical distortion.\cite{kudin_PRB_01}

We did consistency checks for the graphene sphere calculations, three listed next. As a first check, when we investigate Eq.(\ref{eq:criterion}) with graphene parameters, we get $\rho \ll \sqrt{6\text{ \AA}\cdot R}$. For a graphene unit cell $\rho \sim 1$~\AA\ (lattice constant $2.5$~\AA), and the consequent criterion $R\gg 0.2$~\AA\ is easily fulfilled.
We obtained the same $\overline{\kappa}$ with $N=8$ and $N=32$ atom unit cells, even though larger $N$ increases $R_\text{min}$ (Eq.(\ref{eq:criterion}) and $\rho^2 \propto N$ infer $R_\text{min} \propto N$). Thus, the area is small enough to be stress-free, and the simulation is indeed dominated by curvature energy alone. We were able to perform controlled calculations down to radii $R_\text{min} \sim 10$~\AA\, or $\delta \theta_\text{max} \sim 15^\circ$. As a second check, we estimated quantitative error in energy due to the non-commutativity of the two rotations (inset in Fig.\ref{fig:graphene}c), as suggested above, but found the error fairly small. As a third check, we implemented symmetry also with a negative Gaussian curvature $R_1=-R_2=R$, for which $g=-\overline{\kappa}/R^2$ directly, and got an independent confirmation for $\overline{\kappa}$; we won't attempt to describe structures with negative Gaussian curvature here. Finally, since there is no charge transfer, the long-range Coulomb interactions are no issue.


Closer inspection of geometry revealed that curvature increased bond distances as $d_{nn}=1.417$~\AA$+0.135$~\AA $^3/R^2$, due to the weakening of in-plane $\sigma$-bonds, and hereby decreasing the effective nearest-neighbor tight-binding hopping parameter as $t_\text{eff}=t_{gr}-4.8$~eV\AA$^2/R^2$ ($t_{gr}\approx 2.7$~eV). For a detailed discussion of curvature-induced effects on graphene, we recommend Refs.~\onlinecite{kim_EPL_08}, \onlinecite{castro_neto_RMP_09} and \onlinecite{guinea_nphys_10}.

\renewcommand{\arraystretch}{1.2}
\begin{table}[t!]
\caption{Curvature moduli for single- and multi-layer graphene (AB stacking). Numbers in parentheses are estimates from Eq.(\ref{eq:kappas}). $^{a)}$ $\kappa=1.610$~eV for bending against zigzag direction (armchair direction remains straight), and $\kappa=1.606$~eV for bending against armchair direction.}
\begin{tabular}{lcc}
\hline & \\[-10pt]\hline
layers (N) & $\kappa_N$ (eV) & $\overline{\kappa}_N$ (eV) \\
\hline
\vspace{0.1cm}
monolayer & $1.61$ $^a$  & $-0.70$ \\
bilayer  & $180$ ($180$) & $-140$ ($-176$) \\
trilayer & $690$ ($660$) & $-600$ ($-700$) \\
\hline & \\[-10pt]\hline
\label{table}
\end{tabular}
\end{table}

For completeness we calculated $\kappa$ and $\overline{\kappa}$ for bi- and trilayer graphene as well, and summarize the results in Table~\ref{table}. Assuming a constant layer separation of $h=3.4$~\AA\, analytical expressions for the curvature moduli of multi-layer graphene come as
\begin{align}
\begin{split}
\kappa_n = & n\kappa_1+Eh^3(n^3-n)/12 \\
\overline{\kappa}_n= & n\overline{\kappa}_1-Eh^3(n^3-n)/12,
\end{split}
\label{eq:kappas}
\end{align}
where $n$ is the number of layers and $E$ is Young's modulus. The simulated and analytical numbers have a fair agreement. Table reveals how strikingly smaller the moduli are for graphene monolayer, a true oddity among solid elastic sheets, as noted already in Ref.~\onlinecite{yakobson_TAP_01}.

\section{Concluding remarks}

We have introduced a simple and practical method to simulate spherical systems using revised PBC. Although the method is approximate, it is applicable precisely to systems so hard to handle: large systems with prominent curvature effects. Since the method works with schemes from \emph{ab initio} electronic structures and classical potentials to coarse-grained and finite element modeling, and has a wide range of applicability, we encourage any additional implementations.

Admittedly, it may take some time to digest the approximate nature of the method. The role of symmetries in materials modeling is usually taken as clear-cut, solid, and untouchable: it either is or is not. In this paper we have, however, created and entered a new gray area in symmetry usage; we are unaware of symmetry being treated in this type of approximate fashion before. For this reason, when using approximate spherical  symmetry---or other approximate symmetries in future---we urge to examine modeled systems carefully and get assured of method's validity; the best guide on this way is common sense.

\section*{Acknowledgements}

We acknowledge the Academy of Finland for funding, H. H\"akkinen for support, A.~H.~Castro~Neto and T.~Tallinen for inspiring discussions, Jaakko Akola for comments and the Finnish IT Center for Science (CSC) for computational resources.


\begin{thebibliography}{42}
\bibitem[{voi()}]{void}
\bibinfo{note}{With $l$ being the shortest interaction range, the radius of
  void space $R>l/\text{min}(\delta \theta_1,\delta \theta_2)$. Layered
  structure like (void---)water---membrane---water is possible.}
\bibitem[{ele()}]{electrostatics}
\bibinfo{note}{Electrostatics can be refined for spherical symmetry; weighting
  Coulomb interaction by $\sin \theta /\theta$, where $\theta =
  \sqrt{(n_1\delta \theta_1)^2+(n_2\delta \theta_2)^2}$, one gets a
  topologically improved treatment, an approach that should work
  well.\cite{hansen_PRL_79} Instead of Ewald summation that is invalid here, we
  use multipoles for Coulomb summations.}
\bibitem[{Es()}]{Es}
\bibinfo{note}{Estimate is for a thin-sheet calotte with compressive strain
  alone, apart from zero strain at the pole.}
\bibitem[{fma()}]{fmax}
\bibinfo{note}{Spherical approximation forbids optimizing down to zero force,
  which however proved to be a mild restriction. Further, if $f_\text{max}$
  ($0.05$~eV/\AA, say) is a fair criterion for maximum force component for an
  optimized structure in a translational cell, spherical geometry requires a
  tighter criterion $\sim f_\text{max} \times \text{min}(|\delta \theta_1|,
  |\delta \theta_2|)$~eV/\AA.}
\end{thebibliography}
\end{document}